\documentclass{article}
\usepackage{arxiv}
\usepackage{graphicx,amsmath,amsfonts,amscd,amsthm,amssymb,pstricks}
\usepackage{color}
\usepackage{dcolumn}
\usepackage{bm}
\usepackage{longtable}
\usepackage{mathrsfs}
\usepackage{textcomp}
\usepackage{esvect}
\usepackage{float}
\usepackage[USenglish,british]{babel}
\usepackage{environ}
\usepackage{xifthen}
\usepackage{xargs}		
\usepackage{hyperref}
\usepackage{physics,mathtools}
\usepackage[separate-uncertainty = true]{siunitx}
\usepackage{multirow}
\usepackage{comment}

\newcommand{\av}[1]{\left \langle #1\right \rangle }
\newcommand{\ie}{i.e.\ }

\newcommand{\eg}{e.g.\ }

\newcommand{\wrt}{w.r.t.\ }

\newcommand{\ham}{\mathcal{H}}
\newcommand{\eps}{\varepsilon}
\newcommand{\pr}{P}

\def\block(#1,#2)#3{\multicolumn{#2}{c}{\multirow{#1}{*}{$ #3 $}}}
\allowdisplaybreaks[1]

\newtheorem*{theorem*}{Theorem}
\newtheorem*{lemma*}{Lemma}

\title{Nonlinear cooling of an annular beam distribution}

\author{A. Bazzani\\
Dipartimento di Fisica e Astronomia, Universit\`a di Bologna and INFN Bologna, via Irnerio 46, Bologna, Italy\\
\And
F. Capoani\\
Dipartimento di Fisica e Astronomia, Universit\`a di Bologna and INFN Bologna, via Irnerio 46, Bologna, Italy\\
Beams Department, CERN, Esplanade\ des Particules 1, 1211 Geneva 23, Switzerland\\
\And
M. Giovannozzi\thanks{Corresponding author: massimo.giovannozzi@cern.ch}\\
Beams Department, CERN, Esplanade\ des Particules 1, 1211 Geneva 23, Switzerland\\
\And
R. Tom\'as\\
Beams Department, CERN, Esplanade\ des Particules 1, 1211 Geneva 23, Switzerland}

\begin{document}
\maketitle

\begin{abstract}
In recent years, intense efforts have been devoted to studying how nonlinear effects can be used to shape the transverse beam distribution by means of an adiabatic crossing of nonlinear resonances. By this approach, it is possible to split the beams in the transverse plane, so that the initial single-Gaussian beam is divided into several distinct distributions. This is at the heart of the multiturn extraction process that is successfully in operation at the CERN Proton Synchrotron. Nonlinear effects can also be used to cool a beam by acting on its transverse beam distribution. In this paper, we present and discuss the special case of a beam with an annular distribution, showing how its emittance can be effectively reduced by means of properly devised manipulations based on nonlinear effects.
\end{abstract}


\section{Introduction}

Nonlinear effects introduce new beam dynamics phenomena that might open up the possibility of devising novel beam manipulation techniques. This is the case, for instance, when shaping the transverse beam distribution by means of adiabatic crossing of a stable nonlinear resonance. Such a process is at the heart of the so-called beam splitting that is used for the CERN Multiturn Extraction (MTE)~\cite{PhysRevLett.88.104801,PhysRevSTAB.7.024001,PhysRevSTAB.12.014001} and has been successfully implemented as a routine part of operation of the CERN Proton Synchrotron since several years~\cite{Borburgh:2137954,PhysRevAccelBeams.20.014001,PhysRevAccelBeams.20.061001}.

However, this is not the only nonlinear manipulation that can be devised. Indeed, under the inspiration of~\cite{PhysRevLett.110.094801}, it has been found that a controlled redistribution of the invariants can be achieved between the two transverse degrees of freedom~\cite{our_paper7}, provided that an appropriate two-dimensional nonlinear resonance is crossed. This opens novel options in terms of manipulation of the transverse beam emittances. 

It is therefore natural to study whether nonlinear effects can be used efficiently to reduce the linear invariants of a transverse beam distribution, thus generating a cooling of the transverse beam emittance. The basis of this approach to beam cooling is the observation that nonlinear effects do not preserve the linear invariant, \ie the linear action, or the so-called Courant-Snyder invariant. In this sense, they can be used to reduce the value of the linear invariant without violating the symplectic character of the Hamiltonian dynamics. Therefore, the comparison of the value of the linear invariant before and after the action of the nonlinear forces, \ie when the dynamics is linear and expressed as a rotation around the origin of phase space, is a correct indicator of the reduction of the invariant for each individual particle, and hence of the whole beam distribution and of the corresponding emittance. 

In this paper, the initial step towards the development of a nonlinear cooling of a particle distribution is discussed. We present a framework to cool an annular beam distribution, \ie a distribution with nonzero density in an interval of radii $r_1 < r < r_2, \; r_1 > 0$ in the normalized phase space. It is well-known that annular beam distributions are generated as the result of applying a single transverse kick to a centered beam in the presence of decoherence. Hence, a potential application of annular beam cooling could be the restoration of the initial centered distribution after a transverse kick.  

A general discussion of the systems that can be used to devise a cooling method for an annular beam distribution is presented in Section~\ref{sec:gen}, while the considered models are presented in Section~\ref{sec:theory} together with some results of the theory of adiabatic trapping applied to the models. In the same section, several cooling protocols are defined, and their performance analyzed in detail by means of extensive numerical simulations, whose results are presented and discussed in Section~\ref{sec:numres}. Finally, conclusions are drawn in Section~\ref{sec:conc}, with some mathematical details reported in the Appendices.

\section{General considerations on the model chosen} \label{sec:gen}

The general idea underlying the approach developed to achieve cooling of the emittance of an annular beam distribution is based on creating stable islands in phase space. This can be done by slowly modulating the parameters to vary the area of the islands to cause the particles to cross the separatrices. By then moving the resonance islands in phase space their action can be changed and eventually reduced.

To create stable phase-space islands, a resonance needs to be excited. The MTE experience suggests using a H\'enon-like map as a model, close to stable low-order resonances, \eg $1/4$, $1/5$. If the initial annulus lies outside the chain of islands, then by changing the linear frequency one can act on the area of the central region and of the islands to trap particles in the center. This reduces the action by a quantity equivalent to the area of the islands divided by $2\pi$, according to the separatrix crossing theory.

A simple analysis of the scaling laws of the parameters of the islands, found in~\cite{Bazzani:262179}, suggests that this approach is feasible only for resonances of order $n=4$. However, to get the best cooling results one needs two parameters to control the position and the area of the resonance islands. Acting on the sextupolar coefficient is not efficient since this acts as a global-scale parameter~\cite{Bazzani:262179} and hence changes the dynamic aperture of the map. Therefore, an octupolar kick should be added to the sextupolar one to provide an additional free parameter. The estimates for the area of the islands and the central region can be derived using the results of~\cite{Bazzani:262179} and~\cite{PhysRevSTAB.12.024003}. However, the main drawback of this approach is the thick stochastic layer generated by the octupolar kick around the outer part of the separatrix of the four stable islands. This has the effect of inducing the loss of particles, which makes the method unreliable. These observations make the approach based on H\'enon-like maps unsuitable for the application under consideration.

Ongoing studies suggest that trapping into islands and transport from within the islands can also be efficiently achieved using AC-modulated magnets~\cite{our_paper4}. The most straightforward option consists of creating one island using an AC dipole in a $1:1$ resonance condition, \ie with the oscillation frequency close to the linear tune of the system. It is worth recalling that AC dipoles have been widely studied in the field of accelerator physics, with essential applications to beam diagnostics (see \eg~\cite{peggstang,bei1999beam,PhysRevSTAB.5.054001,PhysRevSTAB.8.024401,PhysRevSTAB.11.084002,PhysRevSTAB.16.071002,PhysRevAccelBeams.19.054001,PhysRevAccelBeams.22.031002}, for an overview of AC dipole studies and applications). A cooling method for annular beams will therefore be devised based on Hamiltonian system modeling of the stable islands used to perform the adiabatic trapping, and subsequent transport, under the influence of an AC dipole.

\section{Theory} \label{sec:theory}

\subsection{The Hamiltonian model}

Horizontal betatronic motion in the presence of an AC dipole can be described by the Hamiltonian of a generic oscillator with a sextupolar nonlinearity and a dipolar time-dependent excitation~\cite{PhysRevSTAB.5.054001, peggstang, bei1999beam}, namely
\begin{equation}
\ham(x,p_x,t) = \omega_0 \frac{x^2+p_x^2}{2} + \frac{k_3}{3}x^3 + \eps x \cos\omega t \,  ,
\label{eq:ham_xp}
\end{equation}
where 
\begin{equation}
k_3 = \frac{1}{B_0 \rho} \, \frac{\partial^2 B_y}{\partial x^2} \, \ell \, ,
\end{equation}
and $B_0 \rho$ stands for the magnetic rigidity of the reference particle, $B_y$ is the transverse component of the magnetic field, and $\ell$ is the physical length of the magnetic element. We remark that the choice of the sextupolar nonlinearity is rather arbitrary, as other types of nonlinearity might be used, as long as they generate an amplitude-detuning term. On the other hand, from the standpoint of applications, the use of a sextupolar nonlinearity is very convenient as it is present in all magnetic lattices of circular accelerators.

Using the action angle coordinates $(\phi,J)$ of the unperturbed ($\eps=0$) system and averaging on the fast Fourier components, the Hamiltonian reads
\begin{equation}
\ham(\phi,J,t) = \omega_0 \, J + \frac{\Omega_2}{2} J^2 + \eps\sqrt{2J}\cos\phi\cos\omega t \, ,
\end{equation}
where $\Omega_2 = g(\omega_0) \,k_3^2$ and $g(\omega_0)$ is a function of the linear frequency~\cite{Bazzani:262179}, representing an amplitude tuning term that can be derived using normal forms applied to the Hamiltonian~\eqref{eq:ham_xp}. We recall that $J(x,p_x)$ is an adiabatic invariant of the unperturbed system if the frequency $\omega_0$ is slowly modulated.

If we change the coordinates to refer the system to a rotating reference frame with slow angle $\gamma = \phi-\omega t$, taking into account the generating function $F=J(\phi-\omega t)$ and its time derivative $\pdv{F}{t}=-\omega J$, the transformation gives
\begin{equation}
    \ham(\gamma,J,\psi) = (\omega_0-\omega) J + \frac{\Omega_2}{2} J^2 + \eps\sqrt{2J}\cos(\gamma+\psi)\cos\psi \, ,
\end{equation}
where $\psi=\omega t$.

One can average the fast variable $\psi$, using
\begin{equation}
    \frac{1}{2\pi}\int_0^{2\pi}\dd\psi\,\cos(\gamma+\psi)\cos\psi = \frac{1}{2}\cos\gamma \, , 
\end{equation}
yielding the new averaged Hamiltonian
\begin{equation}
    \ham(\gamma,J) = (\omega_0-\omega) J + \frac{\Omega_2}{2} J^2 + \frac{\eps}{2}\sqrt{2J}\cos\gamma \, ,
\end{equation}
which, after a rescaling, can be written in the following form 
\begin{equation}
    \ham(\gamma,J) = 4J^2 - 2\lambda J + \mu\sqrt{2J}\cos\gamma \, ,
    \label{eq:hamavJ}
\end{equation}
where 
\begin{equation}
\lambda = \frac{4}{\Omega_2}(\omega-\omega_0)\,, \qquad \mu = \frac{4\eps}{\Omega_2} \, .
\label{eq:params}
\end{equation}

Equation~\eqref{eq:hamavJ} represents a well-known Hamiltonian~\cite{neish1975,Neishtadt2013} that can be conveniently written in the form  
\begin{equation}
    \ham(X,Y) = (X^2+Y^2)^2 - \lambda(X^2+Y^2) + \mu X
    \label{eq:hamavxy}
\end{equation}
using the Cartesian coordinates $X=\sqrt{2J}\cos\gamma$, $Y=\sqrt{2J}\sin\gamma$.
When $\lambda > (3/2)\mu^{2/3}$, a hyperbolic fixed point exists only for $Y=0$ and
\begin{equation}
    X = x_\mathrm{c} = \frac{\sqrt{6\lambda}}{3} \cos(\frac{\pi}{6}+\alpha) \, ,
    \label{eq:xc}
\end{equation}
where 
\begin{equation}
    \alpha = \frac{1}{3}\asin(\frac{3\sqrt{6}}{4}\frac{\mu}{\lambda^{3/2}}) \, .
    \label{eq:alpha}
\end{equation}

\begin{figure}[htp]
    \centering
    \includegraphics[width=0.7\columnwidth]{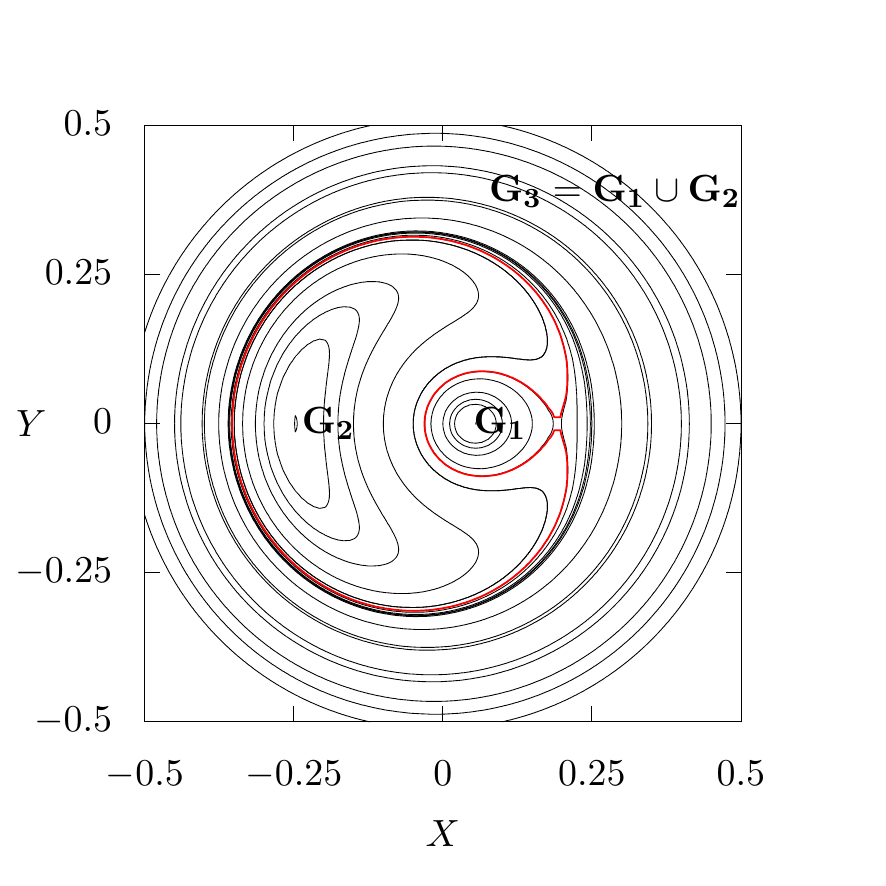}
    \caption{Phase-space portrait of the Hamiltonian~\eqref{eq:hamavxy} with parameters $\lambda=0.1$, $\mu=0.01$. The red line represents the separatrix.}
    \label{fig:ham_phsp}
\end{figure}

The phase space portrait of the Hamiltonian~\eqref{eq:hamavxy} is shown in Fig.~\ref{fig:ham_phsp}, and it can be divided into three regions: the inner regions $G_1$ and $G_2$ (and $G_3=G_1 \cup G_2$) and the region outside them.

Let us compute the area $A_i$ of any region $G_i$. If $\ham_c$ is the value of the Hamiltonian in $(X= x_\mathrm{c}, Y=0)$, the equation $\ham(\gamma, J) = \ham_c$ has the solution
\begin{equation}
    J(\gamma) = \frac{\lambda-2x_\mathrm{c}^2}{2} - 2x_\mathrm{c}\sqrt{\lambda-2x_\mathrm{c}^2}\sin\gamma + 2x_\mathrm{c}^2\sin^2\gamma \, , 
\end{equation}
and $J(\gamma)=0$ for $\gamma=\gamma_0$ with
\begin{equation}
    \gamma_0 = \asin\frac{\sqrt{\lambda-2x_\mathrm{c}^2}}{2x_\mathrm{c}} \, .
\end{equation}
The area of $G_1$ in polar coordinates is thus given by
\begin{equation}
    A_1 = \int_{-\gamma_0}^{\pi - \gamma_0}\dd\gamma\, J(\gamma) = \frac{\pi\lambda}{2}-K_1 - K_2 \, , 
\end{equation}
while the area of $G_3$ is given by 
\begin{equation}
    A_3 = \int_{-\pi-\gamma_0}^{\gamma_0}\dd\gamma\, J(\gamma) = \frac{\pi\lambda}{2}+K_1 + K_2
    \label{eq:A3}
\end{equation}
so that
\begin{equation}
    A_2 = A_3 - A_1 = 2(K_1+K_2) \, , 
\end{equation}
where
\begin{align}
    K_1 &= \lambda\asin(\frac{\sqrt{\lambda-2x_\mathrm{c}^2}}{2x_\mathrm{c}}) \, , \label{eq:K1} \\
    K_2 &= \frac{3}{2}\sqrt{\lambda-2x_\mathrm{c}^2}\sqrt{6x_\mathrm{c}^2-\lambda} \, . \label{eq:K2}
\end{align}

Let us now consider a particle which lies in the outer region with an action $J_0 >  A_3/(2\pi)$. The area enclosed by its orbit will be $A_0 = 2\pi J_0$. If we start a slow modulation of the parameters $\lambda=\lambda(t)$, $\mu=\mu(t)$, according to the theory of adiabatic separatrix crossing~\cite{neish1975,Arnold:937549}, at $t=t^*$, when the condition $A_3=A_0$ is met for $\lambda=\lambda^*$, $\mu=\mu^*$, the particle is captured into $G_1$ or $G_2$ as a random event. Defining
\begin{equation}
    \xi_i = \frac{\dv*{A_i}{t}(\lambda^*,\mu^*)}{\dv*{A_3}{t}(\lambda^*,\mu^*)} \qquad i=1,\,2 \, ,
\end{equation}
the probability $\pr_i$ of trapping in $G_i, \, i=1,\,2 $ is given by
\begin{equation}
\pr_i =\begin{dcases}
\quad 1 &\quad\text{ if } \xi_i > 1\\
\quad\xi_i &\quad\text{ if } 0<\xi_i < 1\\
\quad 0 &\quad\text{ if } \xi_i < 0
\end{dcases}\, .
\end{equation}
After trapping, the resulting action $J$ is given by $A_i/(2\pi)$, where $A_i$ is computed when trapping occurs, namely for $\lambda=\lambda^*$ and $\mu=\mu^*$.

Given an initial distribution of particles, all of which have an initial action in the close neighborhood of $J_0$, the expectation value of their final action is
\begin{equation}
    \av{J}  = {\frac{\pr_1 A_1 + \pr_2 A_2}{2\pi}}\eval_{\lambda^*,\mu^*} \, , 
\end{equation}
and we have $\left \langle J \right \rangle \le J_0$, since $\pr_1+\pr_2=1$, $A_1+A_2=A_3=2\pi J_0$, and $A_i>0$, $\pr_i>0$. Hence, the final expected action is smaller than the initial one, \ie the Courant-Snyder invariant of the particle has been reduced. For a distribution of particles with action $J_0$, this results in a cooling of the beam.

Furthermore, when trapping occurs at ($\lambda^*$, $\mu^*$) we have $A_3 = 2\pi J_0$ and using $A_3 = \pi\lambda^*/2 + K_1 + K_2 = 2\pi J_0$, we obtain the expression
\begin{equation}
    K_1 + K_2 = \pi\qty(2J_0 - \frac{\lambda^*}{2}) \, .
\end{equation}
Substituting $K_1+K_2$ into the expressions for $A_1$ and $A_2$, one obtains
\begin{equation}
    A_1(\lambda^*,\mu^*) = \pi(\lambda^*-2J_0) \qquad A_2(\lambda^*,\mu^*) = \pi(4J_0 - \lambda^*) \, .
\end{equation}

We note that the values of $A_1$ and $A_2$ at the crossing time do not depend on $\mu^*$.

We can then rewrite $\av{J}$ using $\pr_2=1-\pr_1$, which gives
\begin{equation}
    \av{J} = 2J_0-\frac{\lambda^*}{2} + \pr_1(\lambda^* - 3 J_0)
    \label{eq:finJ}
\end{equation}
having calculated $\pr_1$ at $\lambda=\lambda^*$, $\mu=\mu^*$.

\subsection{Cooling protocols} 

We envisage three possible protocols to achieve beam cooling, since we can trap particles by varying only $\lambda(t)$, only $\mu(t)$, or both parameters. We will present the three possible processes in this order, referring to them as Protocol~A,~B and~C, respectively.

\subsubsection{Variation of \texorpdfstring{$\lambda$}{l} (Protocol~A)} \label{sec:theorA}

If we keep $\mu$ constant, $\dv*{A_i}{t} = \pdv*{A_i}{\lambda} \cdot \dv*{\lambda}{t}$, and the probabilities are thus given by
\begin{equation}
    \xi_i = {\frac{\dv*{A_i}{\lambda}(\lambda^*)}{\dv*{A_3}{\lambda}(\lambda^*)}}\eval_{\lambda=\lambda^*} \qquad i=1,\,2 \, .
\end{equation}
Their expressions have been computed in~\cite{neish1975,Neishtadt2013} and read
\begin{equation}
    \begin{aligned}
    \pdv{A_1}{\lambda} &= \frac{\Theta}{2}\,,\qquad &\pdv{A_2}{\lambda} &= \pi - \Theta\,, \\
    \pr_1 &= \frac{\Theta/2}{\pi-\Theta/2}\,, \qquad 
    &\pr_2 &= \frac{\pi-\Theta}{\pi-\Theta/2} \, , 
    \end{aligned}
    \label{eq:probB}
\end{equation}
where
\begin{equation}
    \Theta = \acos(\frac{\lambda}{2x_\mathrm{c}^2}-2) \, .
    \label{eq:theta}
\end{equation}

\begin{figure}[htb]
    \centering
    \includegraphics[width=0.7\columnwidth]{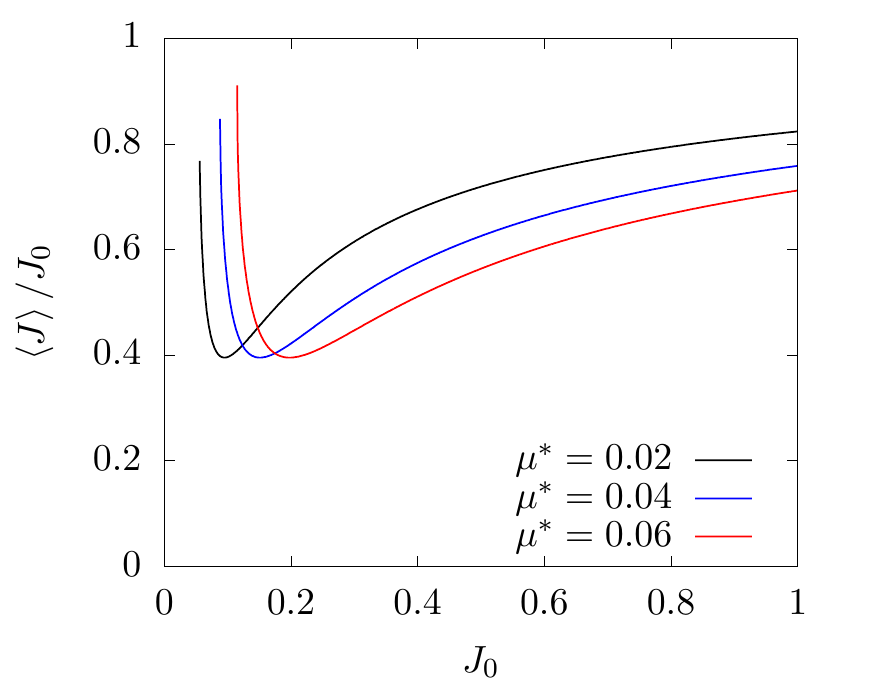}
    \caption{Cooling ratio $\av{J} /J_0$ for trapping in $G_1$ and $G_2$ with the variation of $\lambda$ according to Eq.~\eqref{eq:finJ}, for three values of $\mu^*$.}
    \label{fig:mugraph}
\end{figure}

Figure~\ref{fig:mugraph} shows $\av{J} /J_0$ as a function of $J_0$ for different values of $\mu^*$. We find that the minimum value of $\av{J} /J_0$ is independent of $\mu^*$ (the proof is given in Appendix~\ref{sec:app}).

A numerical computation of this minimum value gives $\av{J} /J_0$ = $\num{0.3957}$. Given $J_0$, we can always find a value $\mu$ that optimizes the cooling, with the final action reduced to $\approx 40\%$ of the initial value.

Strictly speaking, when $\eps\neq 0$, as in the final state of this protocol, the emittance is not equal to the average value of the adiabatic invariant. The reason for this is that the emittance is computed assuming that the dynamics induces a rotation around the origin, whereas the adiabatic invariant is computed with respect to the fixed point around which the initial conditions actually evolve. In fact, when $\eps\neq 0$, and especially when particles are trapped both in $G_1$ and in $G_2$, as in the final state of this protocol, they are not rotating around the origin.  We observe also that if such a cooled beam were transferred to another accelerator, then its emittance would indeed be equal to the average action of the particle distribution. In this sense, the cooling ratio $\av{J} /J_0$ calculated from Eq.~\eqref{eq:finJ} is the lower bound to the actual ratio between the final and initial emittance values.

This situation could be solved or at least mitigated if it were possible to develop a protocol of adiabatic transport that, after the trapping phase, would preserve the actions of the particles while reducing $\mu$ to zero. However, one should consider that when trapping is achieved by means of a variation of $\lambda$ only, the cooling is not particularly efficient, since at best the cooling ratio is $\approx 60\%$. The methods that we are going to present in the following sections are, in theory, capable of achieving total cooling.

\subsubsection{Variation of \texorpdfstring{$\mu$}{m} and  complete trapping in \texorpdfstring{$G_2$}{G2} (Protocol~B)} \label{sec:theorB}

For the protocol based on the variation of $\mu$, the area derivatives $(i=1,\,2)$ are given by 
\begin{equation}
    \dv{A_i}{\mu} = \dv{\alpha}{\mu}\dv{x_\mathrm{c}}{\alpha}\dv{A_i}{x_\mathrm{c}} \, ,  
\end{equation}
where
\begin{equation}
    \begin{aligned}
    \dv{\alpha}{\mu} &= \frac{1}{2}\sqrt\frac{12}{8\lambda^3-27\mu^2}\, , \quad &\dv{x_\mathrm{c}}{\alpha} &= -\frac{\sqrt{6\lambda}}{3}\sin(\frac{\pi}{6}+\alpha)\, , \\
    \dv{A_1}{x_\mathrm{c}} &= -2\frac{(6x_\mathrm{c}^2-\lambda)^{3/2}}{x_\mathrm{c}\sqrt{2x_\mathrm{c}^2-\lambda}}\, , \quad 
    &\dv{A_2}{x_\mathrm{c}} &= 4\frac{(6x_\mathrm{c}^2-\lambda)^{3/2}}{x_\mathrm{c}\sqrt{2x_\mathrm{c}^2-\lambda}} \, .
    \end{aligned}
\end{equation}

Thus, we have $\xi_1 = -1$ and $\xi_2 = 2$, which means that $\pr_1=0$ and $\pr_2=1$. All particles are therefore trapped in $G_2$, with an action value
\begin{equation}
    J = \frac{A_2}{2\pi} = 2J_0 - \frac{\lambda^*}{2}
    \label{eq:theor_m1}
\end{equation}
Cooling is possible in the interval $\lambda^*/4\le J_0 \le \lambda^*/2$, \ie $2J_0\le \lambda^*\le 4J_0$, which corresponds to the existence of the square roots $\sqrt{\lambda^*-2x_\mathrm{c}^2}$ and $\sqrt{6x_\mathrm{c}^2-\lambda^*}$.

On the other hand, for $\lambda^*>4J_0$, the initial condition does not belong to the outer region but to the inner region, $G_1$. In that case, the separatrix crossing occurs when $A_1=2\pi J_0$ and the particle is trapped into $G_2$ at an action $A_2(\lambda^*,\mu^*)/(2\pi)$. Using the expressions of $A_1$ and $A_2$, we find that the resulting expected final action is 
\begin{equation} 
J=\frac{\lambda^*}{2}-2J_0 \, , 
\label{eq:theor_m2}
\end{equation}
which means that cooling is also possible for $4J_0 \le \lambda^* \le 6J_0$, \ie $\lambda^*/6\le J_0 \le \lambda^*/2$.

After being trapped in $G_2$, the particle distribution has a smaller action than the initial one, but, as before, the definition of the adiabatic invariant, being $\mu\neq 0$, is not related to $(x^2+p_x^2)/2$. Therefore, a transport process must be designed to reduce $\mu$ to zero without losing particles from $G_2$. Since the particles are trapped in region $G_2$, we need to keep its area constant, \ie $\dv*{A_2}{t}=0$, or
\begin{equation} 
\dv{A_2}{t} = \dv{\lambda}{t}\qty( \pdv{A_2}{\lambda} + \dv{x_\mathrm{c}}{\lambda} \pdv{A_2}{x_\mathrm{c}} ) = 0 \, .
\label{eq:mutransp}
\end{equation}
This can be used to derive a differential equation for $\mu(\lambda)$
\begin{equation} 
\dv{\mu}{\lambda} = -2x_\mathrm{c}\sqrt\frac{\lambda-2x_\mathrm{c}^2}{6x_\mathrm{c}^2-\lambda}\asin\frac{\sqrt{\lambda-2x_\mathrm{c}^2}}{2x_\mathrm{c}} \, .
\label{eq:odemu}
\end{equation}

Following this equation, as $\lambda$ is reduced $\mu$ increases, and while $A_2$ remains constant $A_1$ is reduced to zero, which occurs when $\mu=(2\lambda/3)^{3/2}$. We can then safely reduce both $\mu$ and $\lambda$ to zero, stopping the perturbation: in fact, as  $\mu$ is kept below $(2\lambda/3)^{3/2}$ no island is present in the phase space.

\subsubsection{Coupled variation of \texorpdfstring{$\lambda$}{l} and \texorpdfstring{$\mu$}{m} and complete trapping in \texorpdfstring{$G_1$}{G1} (Protocol~C)} \label{sec:theorC}

One could also devise a protocol in which both $\lambda$ and $\mu$ are modulated. We can express $\mu$ as a function of $\lambda$, and the expression of the capture probabilities becomes
\begin{equation}
    \pr_i = {\frac{\pdv*{A_i}{\lambda} + \mu' \pdv*{A_i}{\mu}}{\pdv*{A_3}{\lambda} + \mu' \pdv*{A_3}{\mu}}}\eval_{\lambda=\lambda^*,\,\mu=\mu^*} \;\; i=1,\,2 \, ,
\end{equation}
where the prime symbol denotes the derivative \wrt $\lambda$.

\begin{figure*}
    \centering
    \includegraphics[width=\textwidth]{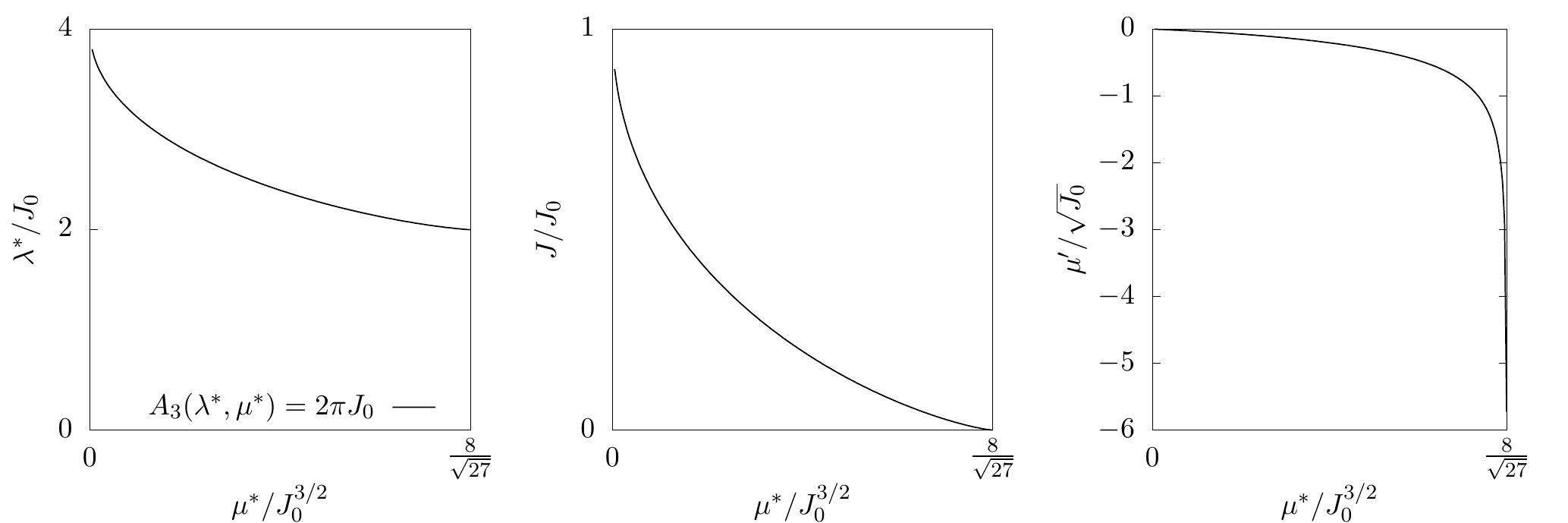}
    \caption{Left: implicit solution $\lambda^*(\mu^*)$ of the equation $A_3(\lambda^*,\mu^*)=2\pi J_0$. Center: expected cooling ratio $J/J_0$ for trapping particles via the coupled variation of $\lambda$ and $\mu$, as a  function of $\mu^*$. Right: required value of $\mu'$ to achieve the cooling efficiency shown in the center plot, as a function of $\mu^*$. Thanks to the ratios of variables reported on the axes, the plotted functions are unique and independent from the value of $J_0$.}
    \label{fig:theor_lm}
\end{figure*}

The trapping probability is calculated at the jumping point $(\lambda^*,\,\mu^*)$. Therefore, we can define an implicit function $\lambda^*(\mu)$ that resolves the equation $A_3 = A_0$ (see Fig.~\ref{fig:theor_lm}, left). Then, we optimize the probability by imposing that: (a) all particles are trapped in region $G_1$; (b) the area $A_1$ is minimized at the trapping point. For the first condition, the equation $\pr_1=1$, $\pr_2=0$, gives the following condition on $\mu'$
\begin{equation}
    \mu' = - {\frac{\pdv*{A_2}{\lambda}}{\pdv*{A_2}{\mu}}}\eval_{\lambda^*,\,\mu^*} \, .
    \label{eq:muprime}
\end{equation}
Note that the signs of the partial derivatives of $A_2$ \wrt $\lambda$ and $\mu$ ensure that $\mu'<0$. 

When $\pr_1=1$, $\pr_2=0$ and $2\pi \av{J}  = A_1 = \lambda^* - 2J_0$, we can minimize $\av{J}$ choosing the minimum $\lambda^*$ for which trapping is possible. This corresponds to $A_1=0$, from which $\lambda^*=2J_0$, and the equation $A_3=2\pi J_0$ becomes
\begin{equation}
    K_1 + K_2 = \pi J_0 \, ,
\end{equation}
that can be solved by setting $K_1=\pi J_0$ and $K_2=0$. From $K_1=\pi J_0$ we have the equation
\begin{equation}
    \asin(\frac{\sqrt{2J_0-2x_\mathrm{c}^2}}{2x_\mathrm{c}}) = \frac{\pi}{2} \, ,
\end{equation}
which is solved when the argument of the arc-sine is $1$, so
\begin{equation}
    \sqrt{2J_0-2x_\mathrm{c}^2} = 2 x_\mathrm{c} \,\implies\, 6x_\mathrm{c}^2 - 2J_0  = 6x_\mathrm{c}^2-\lambda = 0 \, .
\end{equation}

It is straightforward to verify that this implies $K_2=0$. Additionally, this condition induces $\pdv*{A_2}{\mu}=0$, and $\mu'$ diverges. Thus, a perfect cooling, \ie in which the final value of the action is zero, would require to change $\mu$ infinitely fast, which contradicts the adiabatic condition we made to apply the theoretical results. 

Although it is not possible to provide an analytical expression for the implicit solution $\lambda^*(\mu^*)$ of equation $A_3=2\pi J_0$, we can prove that the graphs shown in Fig.~\ref{fig:theor_lm} represent the unique solution after having properly scaled the axes. In particular, we find (the details are reported in Appendix~\ref{sec:app}), that the graph of the implicit solution of equation $A_3=2\pi J_0$ is independent of $J_0$ if we rescale $\lambda^*\to \lambda^*/J_0$ and $\mu^* \to \mu^*/J_0^{3/2}$ (see Fig.~\ref{fig:theor_lm}, left). Similar laws hold for the expected cooling $J/J_0$, which is a function of the only variable $\mu^*/J_0^{3/2}$ (see Fig.~\ref{fig:theor_lm}, center), and for the required $\mu'$, which fulfills the functional relation $\mu'/\sqrt{J_0} = f(\mu^*/J_0^{3/2})$ (see Fig.~\ref{fig:theor_lm}, right).

\section{Simulation results} \label{sec:numres}

We perform numerical simulations of the dynamics generated by the Hamiltonian of Eq.~\eqref{eq:ham_xp} varying $\lambda$ and $\mu$ according to the protocols previously described. In these simulations, we set $\omega_0/(2\pi)=0.414$, $k_3=1$, and invert the relations of Eq.~\eqref{eq:params} to obtain the values of $\omega$ and $\eps$ as a function of $\lambda$ and $\mu$ at each time step. The amplitude-detuning parameter $\Omega_2$ has been evaluated for the unperturbed Hamiltonian at $\eps=0$ by using the algorithm to evaluate the tune described in~\cite{Bartolini:316949}, to give $\Omega_2=-0.3196$.

The initial distributions used in the simulations are an infinitely thin annular distribution with initial action $J_0 = (x_0^2 + p_{x,0}^2)/2$, while uniformly distributed according to the angle variable $\phi_0 = \atan(p_{x,0}/x_0)$, \ie with the p.d.f.\
\begin{equation} 
\rho_{J_0}(\phi, J) = \frac{\delta(J-J_0)}{2\pi} \, .
\end{equation}
%


\subsection{Protocol~A: Cooling by varying \texorpdfstring{$\lambda$}{l}}

This protocol is divided in two phases. The first one is a matching phase, to slowly adapt the initial distribution to the phase space topology, as when $\mu\neq 0$ the elliptic fixed point is shifted. We will increase $\mu$ until the chosen value $\mu^*$ while keeping $\lambda=0$.

In the first phase, for time $t\in [0,t_1]$, we set
\begin{equation}
    \begin{dcases}
    \lambda(t) &= 0\\
    \mu(t) &= \mu^* \frac{t}{t_1} \, .
    \end{dcases}
\end{equation}
The actual trapping occurs in the second phase. The parameter $\lambda$ increases linearly from $0$ to a value $\Delta\lambda$. In order to trap particles at $J_0$, one needs $\Delta\lambda >\lambda^*$, where $\lambda^*=\lambda^*(\mu^*,J_0)$. We then set, for time $t\in [t_1,2t_1]$
\begin{equation}
    \begin{dcases}
    \lambda(t) &= \Delta\lambda \left (\frac{t-t_1}{t_1} \right )  \\
    \mu(t) &= \mu^* \, .
    \end{dcases}
\end{equation}

We remark that although the proposed protocol, for the sake of simplicity, envisages two phases of the same duration, it is certainly possible to remove this constraint to adapt the duration of each phase to make it as adiabatic as possible.


\begin{figure}
    \centering
    \includegraphics[width=0.7\columnwidth]{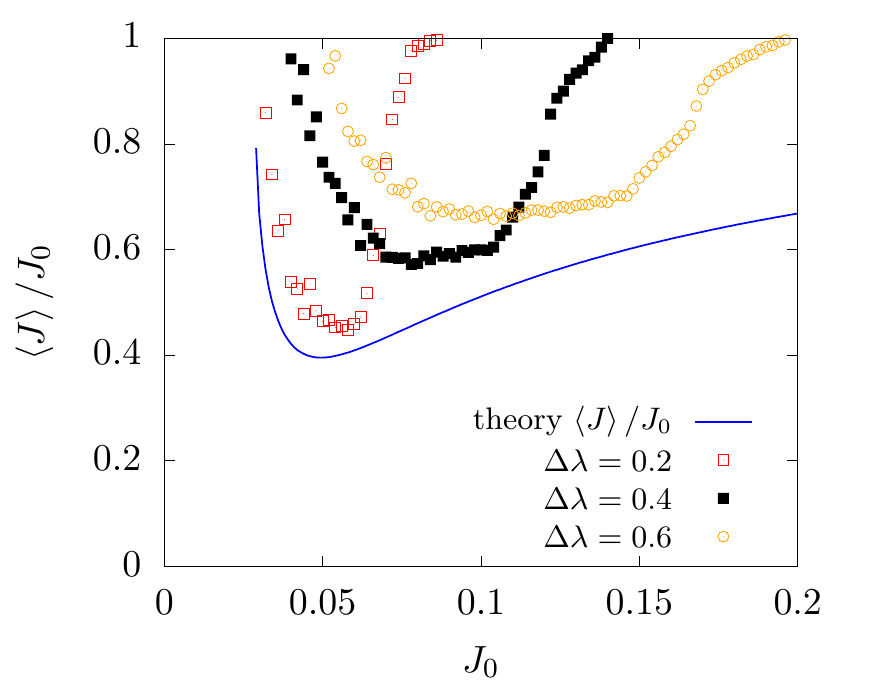}
    \caption{Simulated cooling ratio obtained by applying Protocol~A for different values of $\Delta\lambda$, as function of the initial annular distribution action $J_0$. A comparison with the theoretical bound on the cooling efficiency given by Eq.~\eqref{eq:finJ} is presented. The Hamiltonian~\eqref{eq:ham_xp} has been used, with $k_3=1$, $\omega_0/(2\pi)=0.414$, $\Omega_2=-0.3196$, $\mu^*=\num{7.5e-3}$, $t_1 = \num{5e4}$.}
    \label{fig:result_l}
\end{figure}

Figure~\ref{fig:result_l} shows the simulated $\av{J} /J_0$ for different annular distributions $\rho_{J_0}$ as a function of the initial action $J_0$ using three values of $\Delta\lambda$ (with $\mu^*=\num{7.5e-3}$), and compares it with the theoretical estimate given by Eq.~\eqref{eq:finJ}. We remark that the scale of $\lambda$ and $\mu$ are related with that of $J_0$ and hence the selected values of $\mu^*$ do not have any specific meaning, as any change would simply rescale the $J_0$ axis in Fig.~\ref{fig:result_l}.

We observe two effects that are the root of the difference between the theoretical reduction of $\av{J} /J_0$ and the observed behavior. For larger values of $\Delta\lambda$, the cooling range is increased at the expense of the minimum cooling ratio. Given $\Delta\lambda$ and $\mu^*$, for large values of $J_0$, $\lambda$ is never big enough to achieve trapping, since the $\lambda^*$ value that solves $A_3(\lambda^*,\mu^*)$ is larger than $\Delta\lambda$. Furthermore, increasing $\Delta\lambda$ to trap more particles moves the center of $G_2$ far from the origin of the phase space (all fixed points of Eq.~\eqref{eq:hamavxy}, from the solution of the resulting cubic equation, are $O(\sqrt{\lambda})$ for large values of $\lambda$), thus decreasing the effective cooling ratio. 

\subsection{Protocol~B: Cooling by varying \texorpdfstring{$\mu$}{m}}

This protocol consists of three phases: the first phase is used to perform particle trapping, with the second and the third needed to transport the particles back to the center of the phase space by progressively reducing the strength of the AC dipole. 

In the first phase, for times $t\in[0,t_1]$, we have the following. 
\begin{equation}
    \begin{dcases}
    \lambda(t) & =\lambda^*\\
    \mu(t) &= \mu_1 \frac{t}{t_1} \, ,
    \end{dcases}
\end{equation}
and the condition $\mu_1 > \mu^*$, where $\mu^*$ solves the equation $A_3(\lambda^*,\mu^*)=2\pi J_0$.

In the second phase, the differential equation~\eqref{eq:mutransp}
is solved. For $t\in[t_1,t_2]$, we set $\lambda(t) =\lambda^* - \dot\lambda \left (t-t_1 \right )$ and obtain $\mu(t)$ by numerically integrating the Cauchy problem
\begin{equation}
    \begin{dcases}
    \dv{\mu}{t} &= \dv{\lambda}{t}\dv{\mu}{\lambda} = -\dot\lambda\dv{\mu}{\lambda}\\
    \mu(t_1) &= \mu_1 \, ,
    \end{dcases}
\end{equation}
where $\dv*{\mu}{\lambda}$ is given by Eq.~\eqref{eq:mutransp}. The second phase is stopped at time $t_2$ once the condition $\mu(t_2)=\mu_2 = (2\lambda(t_2)/3)^{3/2}$ is met. The third phase follows for times $t\in [t_2,t_2+t_1]$, with
\begin{equation}
    \begin{dcases}
    \lambda(t) & =\lambda(t_2) \left [1- \left (\frac{t-t_2}{t_1} \right ) \right ]\\
    \mu(t)&= \left ( \frac{2}{3}\lambda(t) \right )^{3/2} \, .
    \end{dcases}
\end{equation}

\begin{figure}[t]
    \centering
    \includegraphics[width=0.7\columnwidth]{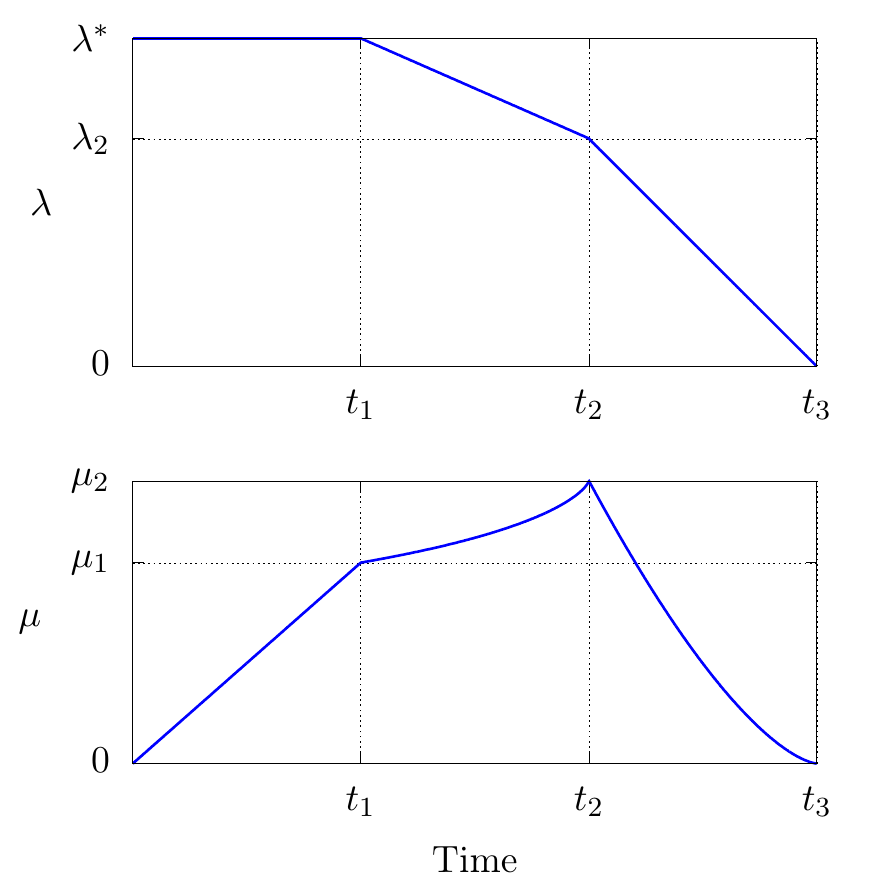}
    \caption{Evolution of $\lambda(t)$ and $\mu(t)$ during the three phases of Protocol~B.}
    \label{fig:protocol_mu}
\end{figure}
The plots of the time evolution of $\lambda$ and $\mu$ are shown in Fig.~\ref{fig:protocol_mu}. 

\begin{figure}[htb]
        \centering
    \includegraphics[width=0.7\columnwidth]{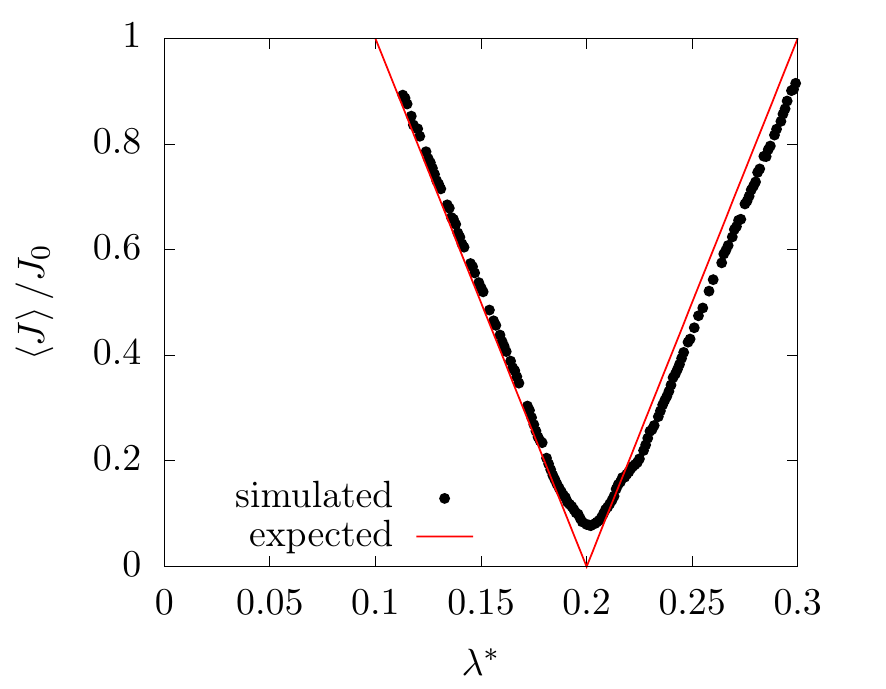}
     \caption{Expected and simulated cooling ratio for trapping in $G_2$ using Protocol~B as a function of $\lambda^*$. The initial distribution is $\rho_{0.05}$. The Hamiltonian~\eqref{eq:ham_xp} has been used, with $k_3=1$, $\omega_0/(2\pi)=0.414$, $\Omega_2=-0.3196$, $\mu_1=0.02$, $t_1 = 1/\dot\lambda = \num{5e4}$.}
    \label{fig:result_m}
\end{figure}

Figure~\ref{fig:result_m} shows the simulated cooling ratio $\av{J}/J_0$, as a function of $\lambda^*$, for an initial annular distribution $\rho_{J_0}(\phi,J)$ with $J_0=\num{0.05}$, together with the theoretical expected value given by Eqs.~\eqref{eq:theor_m1} and~\eqref{eq:theor_m2}.

\begin{figure*}[h]
    \centering
     \includegraphics[trim=5truemm 30truemm 5truemm 5truemm,width=\textwidth,clip=]{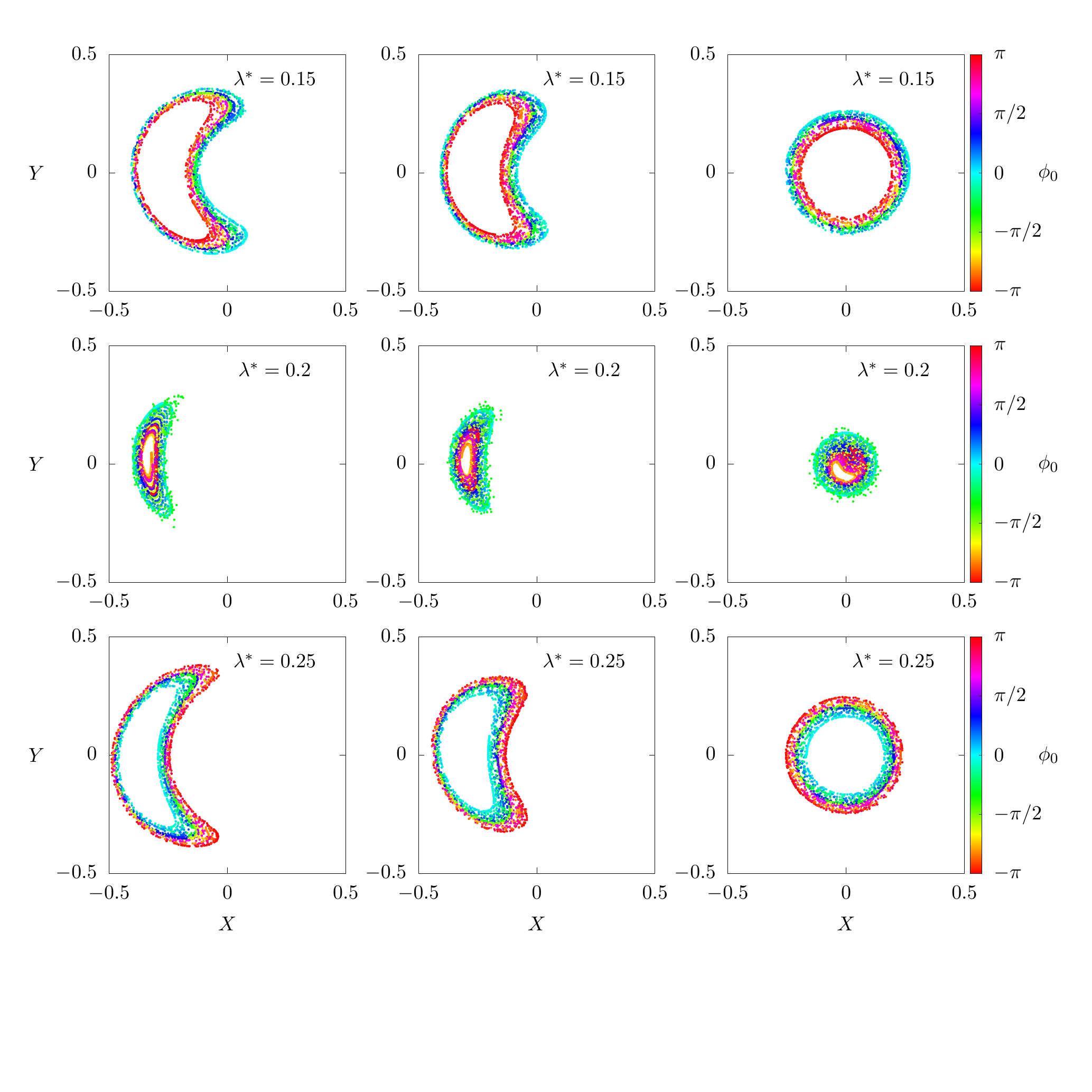}
     \caption{Distributions at the end of the first  (left), second (center), and third (right) phase for an initial distribution $\rho_{0.05}$ for Protocol~B for three values of $\lambda^*$. The color scale represents the initial angle $\phi_0$ and the initial distribution is the same as that shown in the left plots of Fig.~\ref{fig:angle_lm}. Note that for $\lambda^* > 4J_0=0.2$ the angular dependence of the final action is reversed \wrt $\lambda^*<0.2$. The Hamiltonian of Eq.~\eqref{eq:ham_xp} has been used, with $k_3=1$, $\omega_0/(2\pi)=0.414$, $\Omega_2=-0.3196$, $\mu_1=0.02$, $t_1 = 1/\dot\lambda = \num{5e4}$}
    \label{fig:angle_m}
\end{figure*}

We note that the theory presented earlier accurately describes the simulated cooling ratio unless it is in the vicinity of $\lambda^*=4J_0=\num{0.2}$, where the theory predicts total cooling, while in simulation, $\av{J}/J_0\approx 10\%$. This is due to the angular dependence we averaged upon in our analysis, as can be inferred from Fig.~\ref{fig:angle_m}. This figure shows the distributions at the end of each of the three phases of Protocol~B for the same initial annular distribution for three values of $\lambda^*$. We observe that at the end of each phase the action of the particles, which were all the same at the beginning, are spread according to their initial phase. For example red particles, which correspond to the initial phase $\pi$, result in the innermost position when $\lambda^*=0.15$ and in the outermost position when $\lambda=0.25$. This behavior reverses for cyan particles, which have $\phi_0=0$. This means that particles with different initial angles are trapped at slightly different values of $J$. Some particles are trapped earlier or later than expected, with a larger or smaller value of $J$ than that given by theory. In the graphs, it is also visible that the inner and outer particles are reversed, depending on whether $\lambda^*< 4J_0$ or $\lambda^*>4J_0$. When $\lambda\approx 4J_0$, however, all particles are trapped at a higher value than expected no matter when they cross the separatrix, thus increasing $\av{J} $. In our simulations, we were able to reach $\av{J} /J_0 = 0.078$, for a cooling efficiency of $92\%$.
\begin{figure}[htb]
    \centering
    \includegraphics[width=0.7\columnwidth]{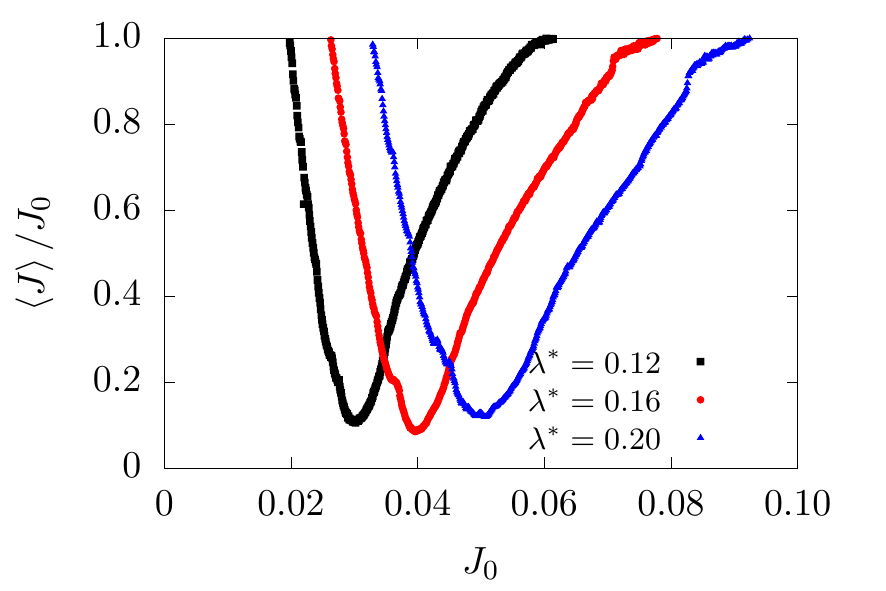}
     \caption{Cooling ratio for trapping in $G_2$ using Protocol~B, at different values of $\lambda^*$, as a function of the initial action of the annular distribution $J_0$. The  Hamiltonian~\eqref{eq:ham_xp} has been used, with $k_3=1$, $\omega_0/(2\pi)=0.414$, $\Omega_2=-0.3196$, $\mu_1=0.02$, $t_1 = 1/\dot\lambda = \num{5e4}$.}
    \label{fig:result_circ_m}
\end{figure}

In Fig.~\ref{fig:result_circ_m} we show the dependence of the cooling ratio on the value of the initial action $J_0$ for three values of $\lambda^*$. The range in which $\av{J} /J_0<1$ represents the possible interval of actions of a thick annular distribution that could be cooled using Protocol~B. Note that according to the theoretical predictions cooling is possible in the range $\lambda^*/6 \leq J_0 \leq \lambda^*/2$ and the optimal cooling ratio is found at $J_0 = \lambda^*/4$.

An animation of the trapping process for a thick annular distribution is available as Supplemental Material~\footnote{See \url{https://gitlab.cern.ch/fcapoani/nonlinear-cooling-animations} for an animation showing the evolution of an initial thick annular distribution under Protocol~B.}

\subsection{Protocol C: Cooling by varying \texorpdfstring{$\lambda$}{l} and \texorpdfstring{$\mu$}{m}}


This protocol requires two phases: the first to adapt the phase space; the second for trapping and transport. Our goal, besides trapping the particles inside $G_1$, is to ensure that both at the beginning and at the end of the process the adiabatic invariant is as close as possible to the linear action variable $J=(x^2+p_x^2)/2$, which is true if the AC dipole is switched off, \ie when $\mu=0$. Thus, in the first phase, $\mu$ is gradually increased, while keeping $\lambda=0$ (\ie $\omega=\omega_0$), until it reaches the value needed to initiate the trapping process. In the second phase, the derivative of $\mu(\lambda)$ is kept at a constant value $\mu'$ while increasing $\lambda$, and taking advantage of the fact that as $\mu'<0$, we can slowly reduce $\mu$ until it reaches zero to recover the equivalence between the adiabatic invariant and $J$.

In the first phase, for times $t\in [0,t_1]$, we set
\begin{equation}
    \begin{dcases}
    \lambda(t) &= 0\\
    \mu(t) &= \mu_\text{max} \frac{t}{t_1} \, , 
    \end{dcases}
\end{equation}
where $\mu_\text{max}=\mu^*+\lambda^*|\mu'|$. This ensures that during the second phase when $\lambda=\lambda^*$, $\mu$ is exactly $\mu^*$ and its derivative $\mu'$ has the appropriate value. The values of $\mu^*$ and $\lambda^*$ are obtained by choosing a solution of the implicit equation $A_3=A_0$ for the selected value of $J_0$ that corresponds to the desired cooling. From Eq.~\eqref{eq:muprime}, the desired value of $\mu'$ is also calculated.

In the second phase, where $t \in [t_1,2t_1]$, we have
\begin{equation}
    \begin{dcases}
    \lambda(t) &=  \frac{\mu_\text{max}}{|\mu'|} \left (\frac{t-t_1}{t_1} \right ) \\
    \mu(t) &=  \mu_\text{max}-|\mu'|\lambda(t) \, .
    \end{dcases} 
\end{equation}

\begin{figure}[htp]
    \centering
    \includegraphics[width=0.7\columnwidth]{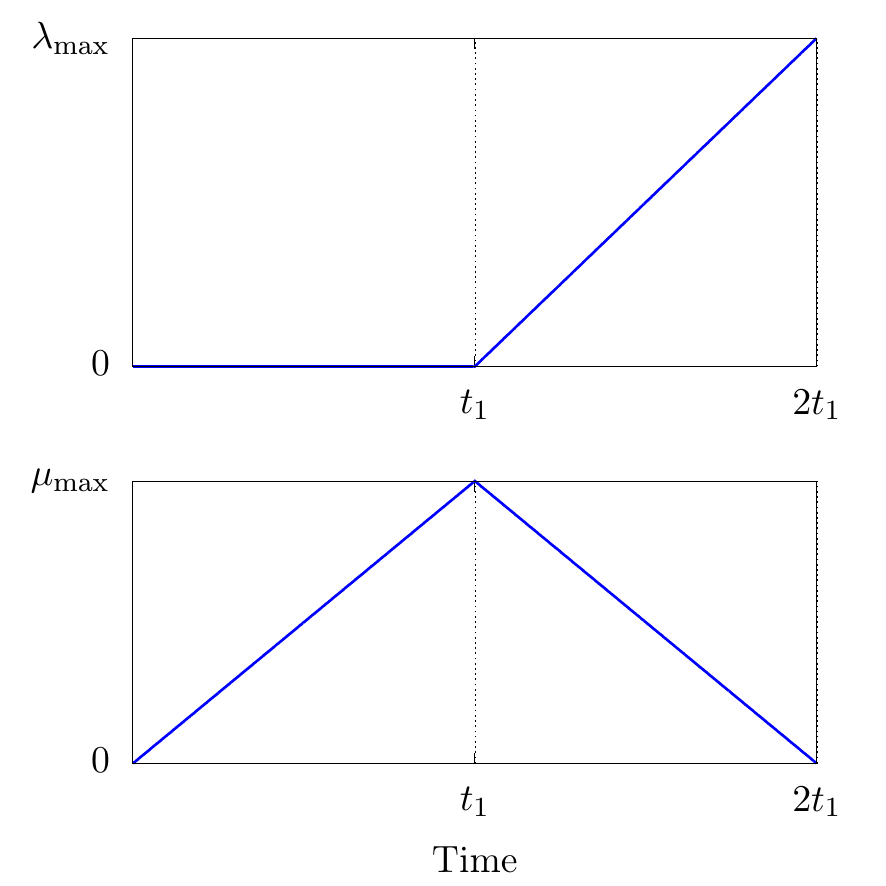}
    \caption{Evolution of $\lambda(t)$ and $\mu(t)$ during the two phases of Protocol~C. The two values $\lambda_\text{max}$ and $\mu_\text{max}$ have expressions in function of the computed $\lambda^*$, $\mu^*$, and $\mu'$, \ie $\mu_\text{max}=\mu^* +\lambda |\mu'|$, $\lambda_\text{max} = \lambda^* + \mu^*/|\mu'|$.}
    \label{fig:protocol_lm}
\end{figure}

When the process ends and $\mu=0$ is reached, $G_2$ disappears as the perturbation provided by the AC dipole has been switched off, and the particles trapped in $G_1$ have been transported to the center of the phase space. The values of $\lambda$ and $\mu$ during the whole procedure are plotted in Fig.~\ref{fig:protocol_lm}. 

We remark that although the proposed protocol envisages two phases of the same duration, it is possible to remove this constraint to adapt the duration of each phase to make them as adiabatic as possible.


\begin{figure}[htb]
    \centering
    \includegraphics[width=0.7\columnwidth]{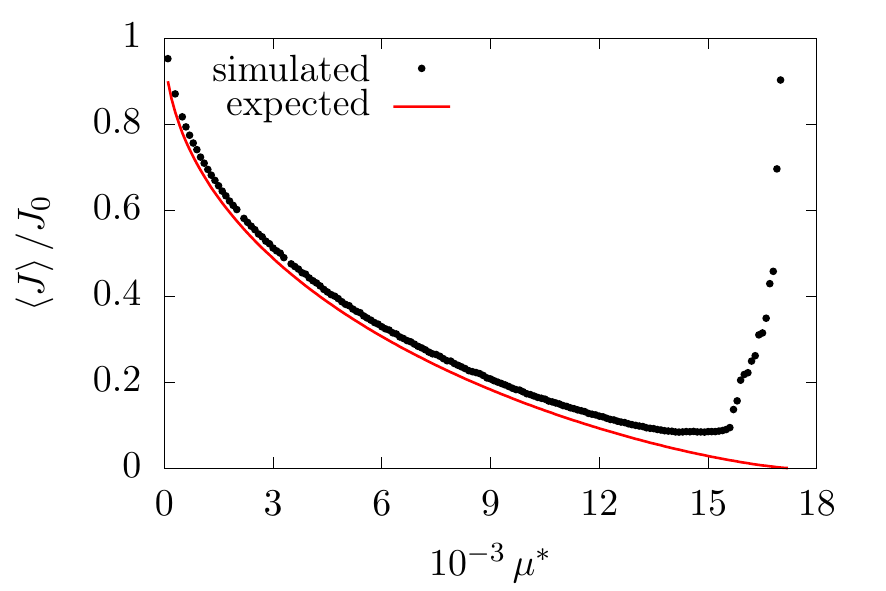}
    \caption{Expected and simulated cooling ratio for trapping in $G_1$ using Protocol~C as a function of $\mu^*$ for an initial distribution $\rho_{0.05}$. The Hamiltonian~\eqref{eq:ham_xp} has been used, with $k_3=1$, $\omega_0/(2\pi)=0.414$, $\Omega_2=-0.3196$, $t_1 = \num{1e5}$.}
    \label{fig:varlm}
\end{figure}

\begin{figure*}[htp]
    \centering
    \includegraphics[trim=10truemm 15truemm 10truemm 0truemm,width=\textwidth,clip=]{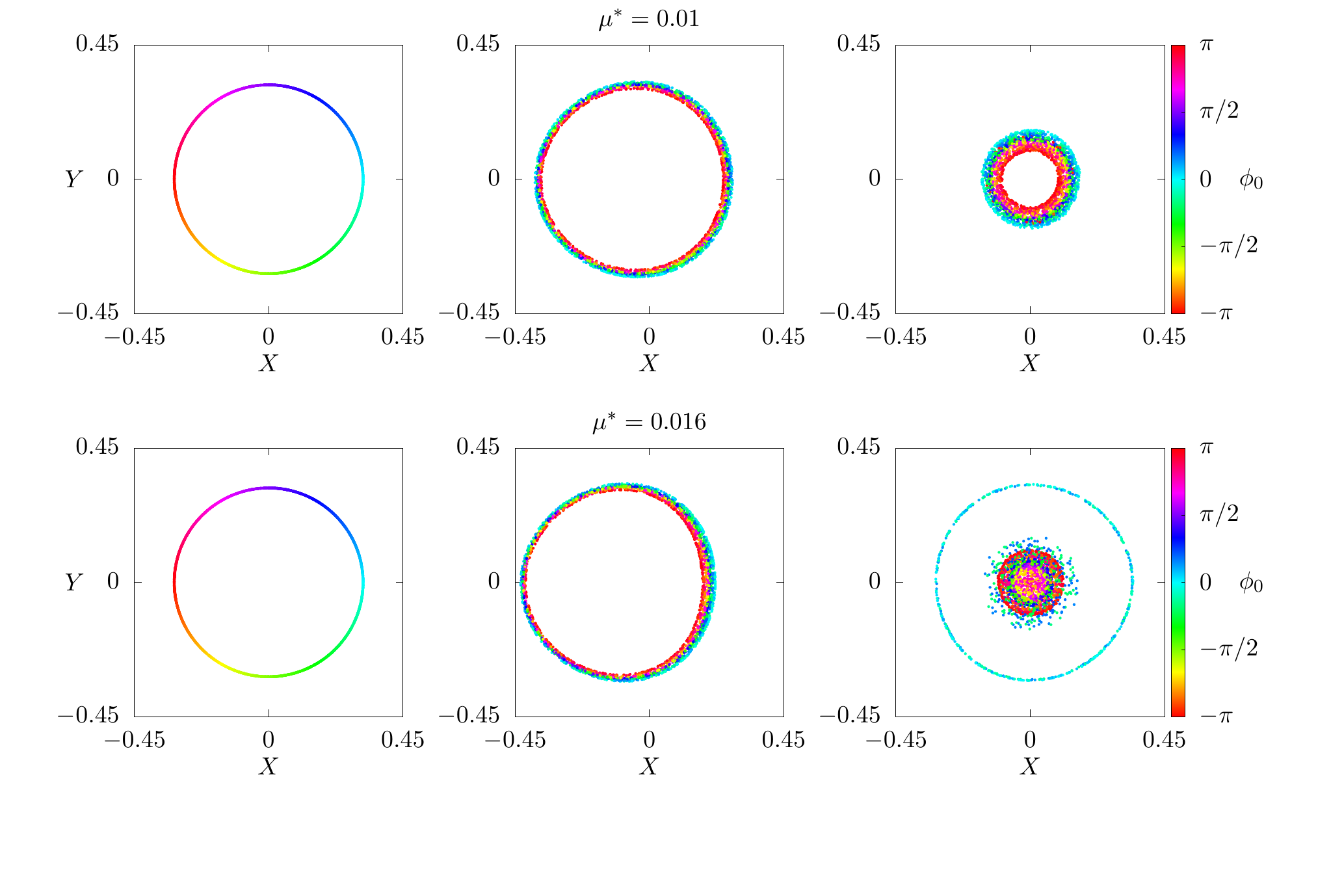}
    \caption{Particle distributions when applying Protocol~C at the beginning (left), after the first phase (middle), and at the end of the second phase (right), for two values of $\mu^*$. The hue encodes the initial angle of the action distribution. The Hamiltonian~\eqref{eq:ham_xp} has been used, with $k_3=1$, $\omega_0/(2\pi)=0.414$, $\Omega_2=-0.3196$, $t_1 = \num{1e5}$.}
    \label{fig:angle_lm}
\end{figure*}

In Fig.~\ref{fig:varlm} we show the simulated cooling ratio $\langle J\rangle/J_0$ for an initial annular distribution $\rho_{0.05}(\phi,J)$, as a function of $\mu^*$, and a comparison with the theoretically expected value $\av{J}  = (\lambda^*(\mu^*) - 2J_0)/(2\pi)$. It can be seen that the agreement between theory and simulation is remarkable up to a certain breakdown value of $\mu^*$. This breakdown is due to the angular dynamics that has been neglected in the averaging process of the theory. In Fig.~\ref{fig:angle_lm} we show the initial distribution, the situation at the end of the first phase and the final distribution of particles for two different values of $\mu^*$, using the hue to represent the initial angle $\phi_0$. For both values of $\mu^*$, we observe that the distribution after the first phase is no longer infinitely thin, and that the action of each particle depends on the initial angle. As a result each particle crosses the separatrix at a different time at the end of the second phase resulting in different values of the final action. For $\mu^*$ smaller than the breakdown threshold, all particles are still trapped in $G_1$, and this angular dependence is averaged out. On the other hand, for higher values of $\mu^*$, particles that at the end of the first phase are in the outer part of the distribution can also be trapped in $G_2$ at high amplitude, thus dramatically increasing the value of the final action. We again stress that we cannot expect to reach $100\%$ cooling as $|\mu'|$ and $\mu_\text{max}$ would need to reach unlimited values. The best cooling that we could achieve in our numerical simulations is $92\%$, at $\av{J} /J_0=0.08$. 
\begin{figure}[htp]
    \centering
    \includegraphics[width=0.7\columnwidth]{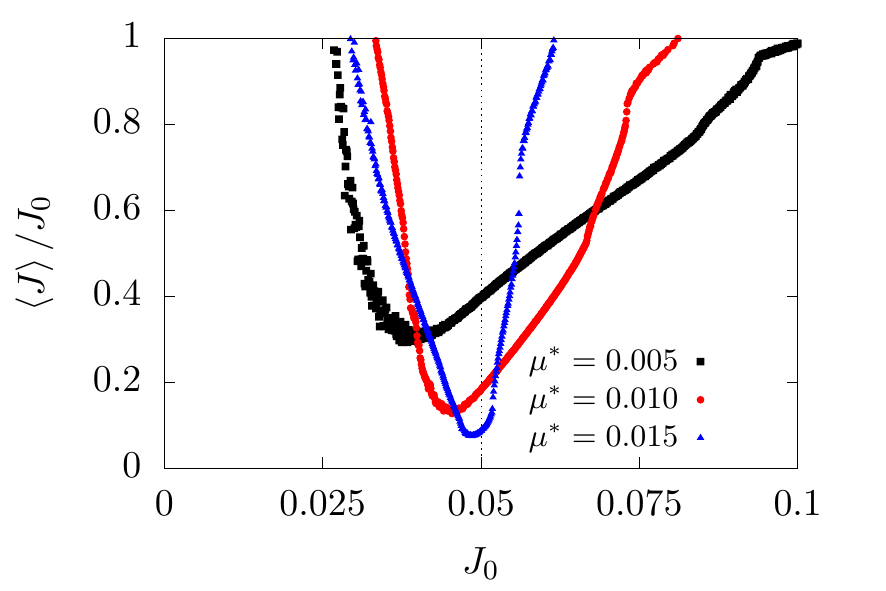}
    \caption{Simulated cooling ratio $\av{J}/J_0$ for initial distributions $\rho_{J_0}$ using Protocol~C at different values of $\mu^*$, having computed $\mu'$ and $\mu_\text{max}$ for $J_0 = \hat J_0 = 0.05$ (indicated by a vertical dotted line in the plot). The Hamiltonian~\eqref{eq:ham_xp} has been used, with $k_3=1$, $\omega_0/(2\pi)=0.414$, $\Omega_2=-0.3196$, $t_1 = \num{1e5}$}
    \label{fig:circles_lm}
\end{figure}

\begin{figure*}[htp]
    \centering
    \includegraphics[trim=5truemm 5truemm 5truemm 0truemm,width=\textwidth,clip=]{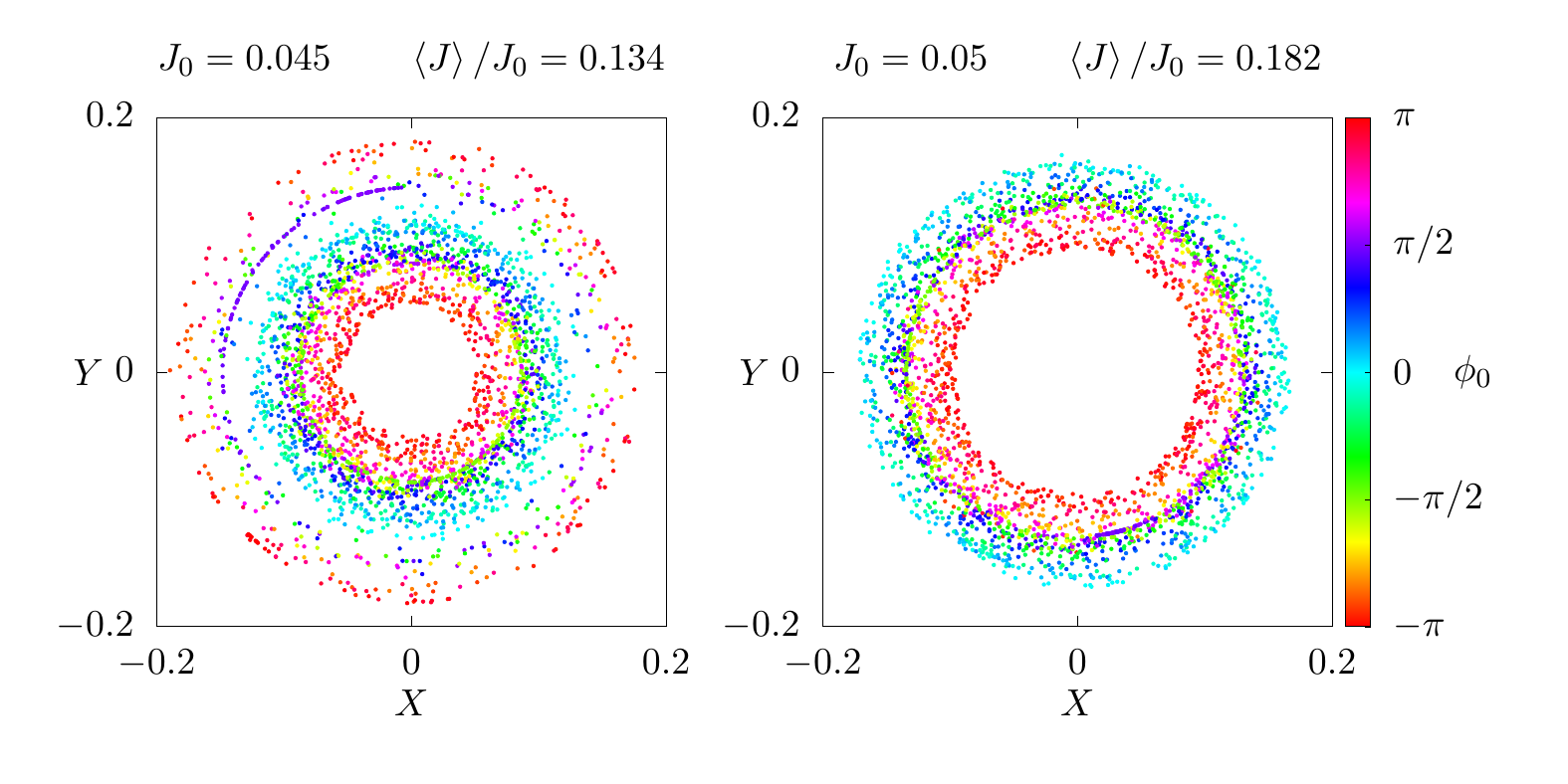}
    \caption{Final particle distributions after applying Protocol~C, for initial distributions $\rho_{0.045}$ (left) and $\rho_{0.05}$ (right), having computed $\mu'$ for $\hat J_0 = 0.05$. The hue indicates the initial angle $\phi_0$. We see that, although the process has been optimized for $\hat J_0=0.05$, for $J_0<\hat J_0$ (and close to the minimum shown in Fig.~\ref{fig:circles_lm}) the resulting cooling is better. The Hamiltonian~\eqref{eq:ham_xp} has been used, with $k_3=1$, $\omega_0/(2\pi)=0.414$, $\Omega_2=-0.3196$, $t_1 = \num{1e5}$, $\mu^*=0.01$.}
    \label{fig:final_example_lm}
\end{figure*}

To study the applicability of the cooling protocol to a more realistic particle distribution, we have looked at an ensemble of infinitely thin annular distributions $\rho_{J_0}$ covering a certain interval in $J_0$. The values of $\mu'$ and $\mu_\text{max}$ have been chosen to optimize the trapping for a particular value of $J_0$, $\hat J_0 = 0.05$. The results are shown in Fig.~\ref{fig:circles_lm}. It is clearly visible that for different values of $\mu^*$, which translates into different cooling targets for particles at $\hat J_0$, a significant range of action values is actually cooled. The width of this \emph{cooling well}, \ie the range of $J_0$ where $\av{J} /J_0<1$, is the thickness of the annular distribution that the protocol can handle successfully. 

We note that, contrary to theoretical expectations, the minimum value of $\av{J} /J_0$ does not occur at $\hat J_0$, although this difference tends to decrease as $\mu^*$ increases. This is due once more to the angular dynamics. Using the same parameters as the plots shown in Fig.~\ref{fig:circles_lm}, two final distributions are shown in Fig.~\ref{fig:final_example_lm} using the hue of the color to identify the initial phase. The right plot shows the case where the initial distribution is $\rho_{0.05}$, \ie the initial conditions are selected at $\hat J_0$, while the left plot shows the case where the initial distribution is $\rho_{0.045}$, where the initial actions have a value $J_0<\hat J_0$, but close to the minimum. In the left plot, a gap in the final distribution is clearly visible. This can be explained by the fact that in this case some particles are trapped earlier (the red dots in the plots) due to the spread of the action after the first phase. These can end up either in $G_2$ or in $G_1$, according to the probability law, but when their areas are smaller. The average final action is therefore reduced more by this effect than by the increase induced by the particles in $G_2$.

An animation of the trapping process for a thick annular distribution is available as Supplemental Material~\footnote{See \url{https://gitlab.cern.ch/fcapoani/nonlinear-cooling-animations} for an animation showing the evolution of an initial thick annular distribution under Protocol~C.}

\section{Conclusions} \label{sec:conc}

In this paper, beam manipulations based on nonlinear beam dynamics have been devised with the goal of achieving cooling for annular transverse beam distributions. Such a beam distribution can be generated after a beam is kicked in the transverse direction. The possibility of achieving cooling by means of crossing stable resonances generated by static magnetic elements has been ruled out, however the use of an AC dipole for such manipulations has proven to be very successful.

A Hamiltonian model describing the transverse dynamics in the presence of an AC dipole has been studied using concepts from the adiabatic theory for Hamiltonian systems. This has allowed the design of three cooling protocols, two of which proved to be extremely effective with a simulated best performance of $\approx 90\%$ cooling. In physical terms this observed cooling is achieved by controlling the strength and frequency of the AC dipole according to the specifications of the proposed protocols.

Detailed numerical simulations carried out on the considered Hamiltonian systems have revealed a rich phenomenology that could be explained in detail by using adiabatic theory for Hamiltonian systems. Although an infinitely thin annular distribution was initially used, the two best protocols have been shown to have a significant cooling range. It therefore seems possible to be able to use them to cool a transverse annular beam distribution of finite thickness. Numerical studies on more realistic accelerator models will be considered in the future in view of experimental tests on a real machine. 

Such annular beam distributions are also representative of the beam halo, which opens up the study of future applications to halo manipulation that could result in experimental tests at the LHC.

\section*{Acknowledgments}

We would like to express our warm thanks to Xavier Buffat for motivating discussions on the topic of this paper. 

\appendix

\section{Some proofs}\label{sec:app} 

Some interesting and useful properties of the theoretical laws that describe the parameters of the cooling protocols described in this paper can be derived by reasoning on the functional dependencies. Note that in the following, $f(x)$, $g(x), A(x), B(x)$ etc.\ represent generic functions of the only variable $x$, and the same occurs for their product, \ie $f(x)g(x)=h(x)$.

\subsection{Uniqueness of the minimum of \texorpdfstring{$\av{J} /J_0$}{<J>/J0} for Protocol~A}

From the expressions of $x_\mathrm{c}$, $K_1$, $K_2$ and $\Theta$ (see Eqs.~\eqref{eq:xc},~\eqref{eq:alpha},~\eqref{eq:K1},~\eqref{eq:K2},~\eqref{eq:theta}), we define $\chi = \mu/\lambda^{3/2}$, and we can express these quantities as $x_\mathrm{c} = \sqrt\lambda \tilde{x}_\mathrm{c}(\chi), K_i=\lambda\tilde{K}_i(\chi)$, and $\Theta = \Theta(\chi)$.

From the relation $A_3=2\pi J_0$, we have $J_0 = \lambda^* f(\chi^*)$, while from Eq.~\eqref{eq:finJ} one finds that $\av{J}  = \lambda^* g(\chi^*)$. Hence, setting $h(\chi)=g(\chi)/f(\chi)$, we finally have $\av{J} /J_0 = h(\chi^*)$. From the expression of $f(\chi^*)=\av{J}/\lambda^*$, and noting that $g(\chi^*)=J_0/\lambda^*$ is monotone (see Fig.~\ref{fig:theor_lm}, left) it is possible to show that the function $h(\chi^*)$ has a minimum for a value $\hat{\chi}^*$.

Then, there exists only one pair $(\lambda^*,\,\mu^*)$ that solves $A_3=2\pi J_0$ and for which $\mu^*/\lambda^{*3/2} = \hat{\chi}^*$. Therefore, for each $J_0$ there exists only one value $\hat{\chi}^*$ and, therefore, a unique value of $h(\hat{\chi}^*)$, which does not depend on $\mu^*$. This proves what has been observed in Section~\ref{sec:theorA}.

\subsection{Scaling laws for Protocol~C}

A similar approach can be used to derive the scaling laws of Section~\ref{sec:theorC}. As $A_3 = \lambda^* \tilde{A}_3(\chi^*)$, the equation defining the invariant after the trapping reads
\begin{equation} 
\lambda^*\, \tilde A_3(\chi^*) = \lambda^*\, \tilde A_3\qty(\frac{\mu^*}{{\lambda^*}^{3/2}}) = 2\pi J_0 \, .
\label{eq:funcJ0}
\end{equation}

The functional equation
\begin{equation}
    x\, f\qty(\frac{y}{x^\alpha}) = 2\pi z
\end{equation}
under the transformations $\overline x=x/z$, $\overline y = y/z^\alpha$ becomes
\begin{equation}
    \overline x\, f\qty(\frac{\overline y}{\overline x^\alpha}) = 2\pi \, ,
\end{equation}
and this implicit equation is solved by the function $\overline x=g(\overline y)$, whence we infer that, after rescaling $\lambda^*\to \lambda^*/J_0$ and $\mu^* \to \mu^*/J_0^{3/2}$, the function
\begin{equation}
\frac{\lambda^*}{J_0} = A\qty(\frac{\mu^*}{J_0^{3/2}})
\label{eq:funcA}
\end{equation}
represents the unique solution to Eq.~\eqref{eq:funcJ0}. This explains the scaling shown in Fig.~\ref{fig:theor_lm} (left).

Moreover, inverting Eq.~\ref{eq:funcJ0} one finds that $\chi^*$ can be written as a function of $\lambda^*/J_0$, and therefore of $\mu^*/J_0^{3/2}$:
\begin{equation}
\chi^* = \chi^*\qty(\frac{\lambda^*}{J_0}) = \chi^*\qty(A\qty(\frac{\mu^*}{J_0^{3/2}})) = \chi^*\qty(\frac{\mu^*}{J_0^{3/2}})\, .
\label{eq:funcChi}
\end{equation}

We can therefore find a scaling law for the expected cooling ratio, as $ 2 \pi J = A_1 = \lambda^* \tilde A_1 (\chi^*)$, and using Eq.~\eqref{eq:funcA} and Eq.~\eqref{eq:funcChi} one obtains
\begin{equation}
    \begin{split}
    \frac{J}{J_0} &= \frac{\lambda^*}{J_0}\tilde J(\chi^*) = A\qty(\frac{\mu^*}{J_0^{3/2}}) \tilde J\qty(\chi\qty(\frac{\mu^*}{J_0^{3/2}})) = \\ & =B\qty(\frac{\mu^*}{J_0^{3/2}}) \, ,
    \end{split}
    \label{eq:theor_lm}
\end{equation}
which explains the scaling shown in Fig.~\ref{fig:theor_lm} (center).

In the same way, the coefficient $\mu'$, from the expressions of $\pdv*{A_2}{\lambda}$ and $\pdv*{A_2}{\mu}$ in Eqs.~\eqref{eq:probB} and~\eqref{eq:theta}, can be written as
\begin{equation}
    \mu' = \sqrt{\lambda}\,\tilde \mu'\qty(\chi) \, , 
\end{equation}
and, dividing by $\sqrt{J_0}$ and using the functional dependencies of Eqs.~\eqref{eq:funcA} and~\eqref{eq:funcChi} one obtains
\begin{equation}
    \frac{\mu'}{\sqrt J_0} = \sqrt{\frac{\lambda^*}{J_0}}\tilde\mu'\qty(\chi^*) =C\qty(\frac{\mu^*}{J_0^{3/2}}) \, , 
\end{equation} 
which is the scaling for the plot shown in Fig.~\ref{fig:theor_lm} (right).

\bibliographystyle{unsrt}
\bibliography{mybibliography}
\end{document}